\documentclass[prl,twocolumn,showpacs,nofootinbib,aps,longbibliography,superscriptaddress]{revtex4-1}
\usepackage[utf8]{inputenc}
\usepackage{color}
\usepackage{amsmath}
\usepackage{url}
\usepackage{graphicx}
\usepackage{amssymb}
\usepackage[dvipsnames]{xcolor}
\usepackage{appendix}
\usepackage[section]{placeins}
\usepackage{multirow}
\usepackage{hyperref}
\usepackage{bm}
\usepackage{soul}
\usepackage{array}
\usepackage{lipsum}
\usepackage{relsize}
\usepackage{amsfonts}
\usepackage{comment}
\usepackage[normalem]{ulem}
 \usepackage{float}

\DeclareMathOperator*{\argmax}{arg\,max}

\usepackage[ruled,lined]{algorithm2e}

\newcommand{\enm}{E^\text{NM}}
\newcommand{\rhoc}{\rho_\text{c}}
\newcommand{\knm}{K^\text{NM}}

\newcommand{\asym}{a_\text{sym}^\text{NM}}
\newcommand{\lsym}{L_\text{sym}^\text{NM}}

\newcolumntype{C}[1]{>{\centering\arraybackslash}p{#1}}

\begin{document}

\title{Training and Projecting: A Reduced Basis\texorpdfstring{\\}{ }Method Emulator for Many-Body Physics}


\author{Edgard Bonilla}
\altaffiliation{Bonilla and Giuliani are co-authors contributing equally to this work}
\email[\\]{edgard@stanford.edu}
\affiliation{Department of Physics, Stanford University, Stanford, California 94305, USA}
\author{Pablo Giuliani}

\email{giulianp@frib.msu.edu}
\affiliation{FRIB/NSCL Laboratory, Michigan State University, East Lansing, Michigan 48824, USA}
\affiliation{Department of Statistics and Probability, Michigan State University, East Lansing, Michigan 48824, USA}

\author{Kyle Godbey}
\email{godbey@frib.msu.edu}
\affiliation{FRIB/NSCL Laboratory, Michigan State University, East Lansing, Michigan 48824, USA}

\author{Dean Lee}
\email{leed@frib.msu.edu}
\affiliation{FRIB/NSCL Laboratory, Michigan State University, East Lansing, Michigan 48824, USA}
\affiliation{Department of Physics and Astronomy, Michigan State University, East Lansing, Michigan 48824, USA}

\date{\today}

\begin{abstract}

We present the reduced basis method as a tool for developing emulators for equations with tunable parameters within the context of the nuclear many-body problem. The method uses a basis expansion informed by a set of solutions for a few values of the model parameters and then projects the equations over a well-chosen low-dimensional subspace. We connect some of the results in the eigenvector continuation literature to the formalism of reduced basis methods and show how these methods can be applied to a broad set of problems. As we illustrate, the possible success of the formalism on such problems can be diagnosed beforehand by a principal component analysis. We apply the reduced basis method to the one-dimensional Gross-Pitaevskii equation with a harmonic trapping potential and to nuclear density functional theory for $^{48}$Ca, achieving speed-ups of more than x150 in both cases when compared to traditional solvers. The outstanding performance of the approach, together with its straightforward implementation, show promise for its application to the emulation of computationally demanding calculations, including uncertainty quantification.
\end{abstract}
\maketitle

Most modern theoretical models describing many-body nuclear dynamics share an ever-increasing computational burden. This can turn into a challenge tasks like uncertainty quantification (UQ) analysis~\cite{phillips2021get,ekstrom2019bayesian}, experimental design~\cite{melendez2021designing,giuliani2021noise}, calibration of model parameters~\cite{pratt2015constraining,mcdonnell2015uncertainty,wesolowski2019exploring}, and even repeated evaluation for different inputs~\cite{lasseri2020taming,godbey2022}. Emulators---algorithms able to give fast-and-accurate approximate calculations to expensive computations---have been gaining increasing importance as a way to circumvent these challenges~\cite{santner2003design,phillips2021get}.

In recent years, a technique called eigenvector continuation (EC)~\cite{frame2018eigenvector} was developed to emulate computationally intensive calculations involving bound states of Hamiltonian operators ~\cite{konig2020eigenvector} and nuclear scattering~\cite{furnstahl2020efficient,melendez2021fast,drischler2021toward}. EC has shown excellent performance in interpolation and extrapolation by working with two elements: choosing its ansatz functions from the linear span of exact solutions to the problem at hand, and using a variational principle---for example the Rayleigh-Ritz method~\cite{frame2018eigenvector,cohen1986quantum} or the Kohn variational principle~\cite{furnstahl2020efficient,kohn1948variational}---to obtain equations for the coefficients of this linear combination.

We present an emulator constructed in the formalism of reduced basis methods (RBMs) ~\cite{almroth1978automatic,quarteroni2015reduced,hesthaven2016certified}, a set of dimensionality-reduction techniques that fall under the umbrella of reduced order models \cite{quarteroni2014reduced,brunton2019data,melendez2022model}. These methods have seen active development over the last two decades, proving to be useful in a variety of computationally-intensive problems involving partial differential equations ~\cite{quarteroni2011certified,nguyen2010reduced,field2011reduced,MILANI20084812}. EC can be naturally connected to RBMs by constructing a generalization of EC through a Galerkin method formulation. The key insight is that once a reasonable choice of ansatz functions has been made (for example, the EC basis), all that is needed is a method to select a suitable candidate approximation from the ansatz subspace. This could be achieved, for instance, by using a variational principle, minimizing a cost functional, or by finding the fixed point of an iterative scheme. Among the alternatives, the Galerkin method---the option chosen in RBMs---stands out for its simplicity: it attempts to find an accurate approximate solution by projecting the problem to a well-chosen very-low dimensional subspace. This simplicity allows these methods to be applied to a wide variety of problems in a straightforward way.

RBMs are tailored to problems that feature an equation that depends smoothly on a list of tunable control parameters $\alpha$~\cite{quarteroni2015reduced}. The goal is to build an approximate solution for a suitable range of these parameters. Let us assume the equation is written in the general form:
\begin{equation}
    F_\alpha(\phi_\alpha)=0, \label{eqn:non_linear1}
\end{equation}
where $\phi_\alpha$ is a vector (or function) from a Hilbert space $\mathcal{H}$, and $F_\alpha$ maps $\mathcal{H}$ onto itself. For example, in the case of bound systems with a Hamiltonian $H_\alpha$ that depends on $\alpha$, $F_\alpha$ can take the form of the eigenvalue equation $F_{\alpha}(\phi_{\alpha})=H_{\alpha}\phi_{\alpha}-\lambda_{\alpha} \phi_{\alpha}$, where $\lambda_\alpha$ is the eigenvalue. Another example would be the case of single-channel scattering where $F_\alpha$ can be the radial part of the scattering equation~\cite{thompson2009nuclear} $ F_\alpha(\phi_\alpha)=\left(-\frac{d^2}{dr^2}+\frac{\ell(\ell+1)}{r^2}+U(r,\alpha)-p^2\right)\phi_\alpha(r)$, where a system with reduced mass $\mu$ interacts through a potential $V(r,\alpha)=U(r,\alpha)/2\mu$ with parameters $\alpha$, $\ell$ is the angular momentum quantum number, and $p$ is the asymptotic linear momentum. The RBM finds approximate solutions $\hat{\phi}_\alpha$ to these---and more general---problems by constructing a basis expansion with $n$ linearly independent `reduced basis' functions $\{\phi_k\}_{k=1}^n$:
\begin{equation}
    \hat{\phi}_\alpha=\phi_0+\sum_{k=1}^na_k\phi_k,\label{eqn:training_f}
\end{equation}
where $\phi_0$ is an extra term that can be added to satisfy boundary conditions imposed on Eq.~\eqref{eqn:non_linear1}. The reduced basis functions $\phi_k$ are selected to create an affine space (the ansatz subspace) \emph{close} to the manifold formed by the solutions to Eq.~\eqref{eqn:non_linear1} as a function of the parameters $\alpha$~\cite{quarteroni2015reduced} by using the information from a (possibly small) sample of exact solutions. In practice, the `exact solutions' (or `snapshots') of Eq.~\eqref{eqn:non_linear1} are constructed by highly accurate yet computationally expensive approximations such as finite element or spectral calculations~\cite{quarteroni2011certified}.
 
\begin{figure}[t]
    \centering
    \includegraphics[width=0.48\textwidth]{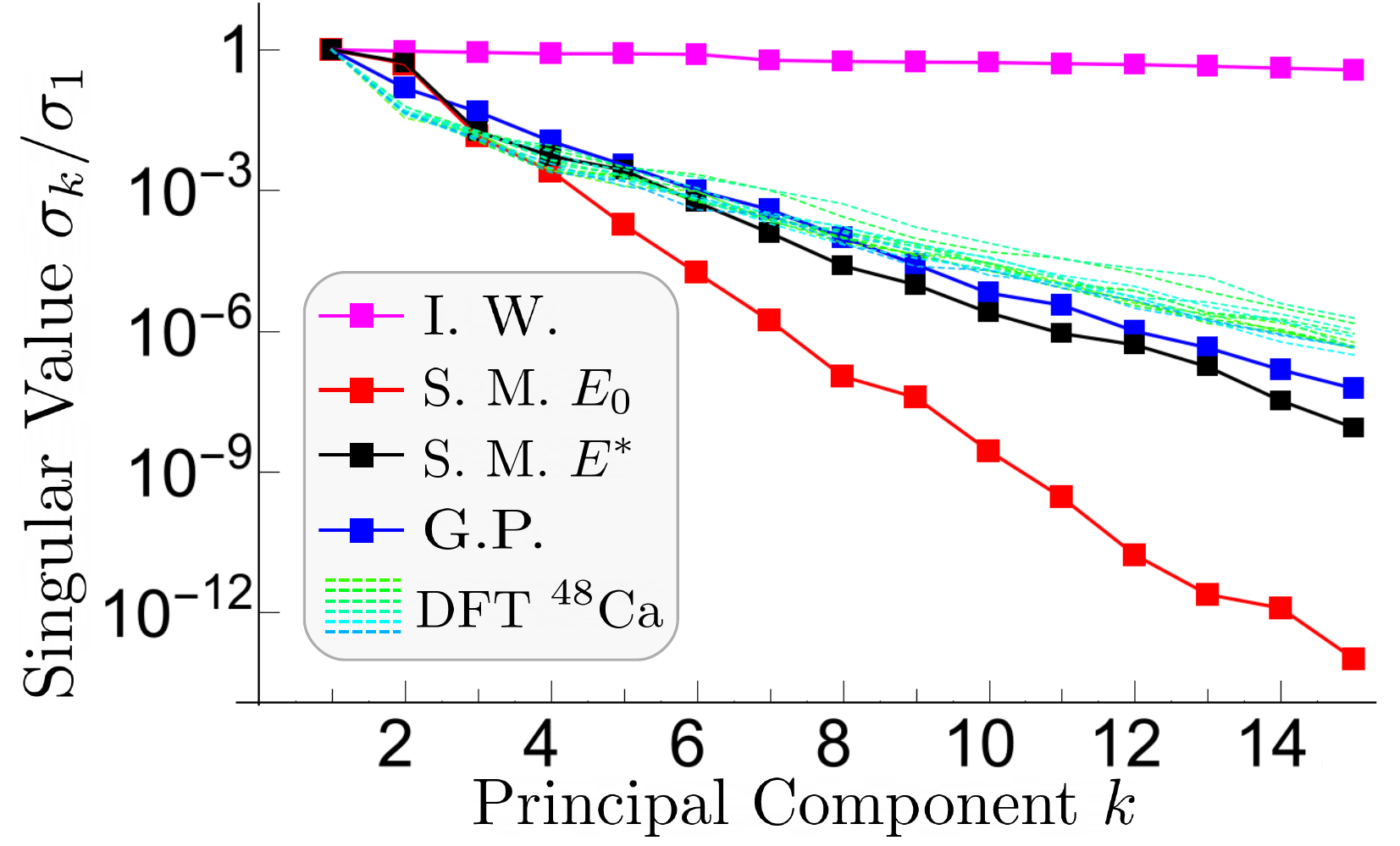}
    \caption{Decay of the singular values $\sigma_k$ for a set of solutions of: the Infinite Well (I.W. in magenta) with $V(x,\alpha)=0$ for $x\in[\alpha,\alpha+1]$ and $V(x,\alpha)=\infty$ otherwise, the single channel 2-body Scattering with a Minnesota potential~\cite{thompson1977systematic} with fixed energy (S.M. $E_0=50$ MeV in red) and varying energy (S.M. $E^{*}\in[20,80]$ MeV in black) with the parameter's range as used in Ref.~\cite{furnstahl2020efficient}, the Gross-Pitaevskii equation (G.P. in blue), and the solutions for the 13 energy levels in $^{48}$Ca (DFT in dashed green-blue lines). The supplementary material contains additional details on these calculations, including the ranges for the values of $\alpha$ used.}

    \label{fig:SVD}
\end{figure}

An option for building the reduced basis in Eq.~\eqref{eqn:training_f} is the approach taken in many EC applications~\cite{frame2018eigenvector,konig2020eigenvector,furnstahl2020efficient,melendez2021fast,drischler2021toward}, also known as the Lagrange basis~\cite{quarteroni2011certified}. It consists of calculating $n$ `training functions' $\{\tilde\phi_k\}^n_{k=1}$ as solutions to Eq.~\eqref{eqn:non_linear1} for $n$ values $\alpha_k$ ($F_{\alpha_k}(\tilde\phi{_k})=0$), and then choosing the reduced basis as these $n$ training functions $\phi_k=\tilde\phi_k$. One possible way to improve upon this choice is the so-called proper orthogonal decomposition (POD)~\cite{quarteroni2015reduced}. It consists of computing $N\geq n$ solutions $\{\tilde\phi_l\}^N_{l=1}$ and constructing the reduced basis with the first $n$ components from a principal component analysis (PCA)~\cite{jolliffe2002principal}, or singular value decomposition (SVD)~\cite{blum_hopcroft_kannan_2020}, of the set of these $N$ training functions. Therefore, by using the information of $N$ samples, the POD basis is more robust than a Lagrange basis of dimension $n$, and faster than a Lagrange basis of $N$ training points. As a side note, an exponential decay on the associated singular values $\sigma_k$ from the PCA indicates that the RBM can provide an accurate approximation for the problem at hand~\cite{quarteroni2011certified,quarteroni2015reduced,nonino2019overcoming}. We exploit this feature later when discussing Fig.~\ref{fig:SVD}.

Once the $n$ reduced basis functions are chosen, the coefficients $a_k$ for the approximation are found by the Galerkin method~\cite{rawitscher2018galerkin}, that is, by projecting Eq.~\eqref{eqn:non_linear1} over $n$ linearly independent `projecting functions' $\{\psi_j\}_{j=1}^n$ in the Hilbert space:
\begin{equation}
    \langle \psi_j|F_\alpha(\hat{\phi}_\alpha)\rangle=0,\quad \text{for all }j.\label{eqn:projections}
\end{equation}

$F_\alpha(\hat{\phi}_\alpha)$ is often called the residual~\cite{fletcher1984computational}, and it can be used, for example, to inform the construction of the reduced basis~\cite{buffa2012priori}, or to estimate the emulation error~\cite{Prudhomme2001Reliable}. We can interpret Eq.~\eqref{eqn:projections} as enforcing the orthogonality of $F_\alpha(\hat\phi_\alpha)$ to the subspace spanned by $\{\psi_j\}_{j=1}^n$, i.e., by finding a $\hat\phi_\alpha$ such that $F_\alpha(\hat\phi_\alpha)$ is ``zero'' up to the ability of the set $\{\psi_j\}_{j=1}^n$. The choice of projecting functions $\psi_j$ is arbitrary, but is usually also informed by the solution manifold~\cite{quarteroni2015reduced,hesthaven2016certified}. For the rest of this work, we choose $\psi_j$ to enforce orthogonality with respect to the ansatz subspace \eqref{eqn:training_f}, which is the traditional way of using the Galerkin method~\cite{fletcher1984computational}. 

The reduced-basis emulators are most effective, in terms of speed ups, when the projections in Eqs.~\eqref{eqn:projections} lead, for every $j$, to expressions of the form:
 \begin{equation}
     \sum_{m=1}^{M_j}f_{j,m}(\alpha)g_{j,m}(a_1,\dots,a_n)=0, \label{eqn:affine_decomposition}
 \end{equation}
where $f_{j,m}(\alpha)$ and $g_{j,m}(a_1,...a_n)$ are $M_j$ functions that are independent of the intrinsic coordinates of the original system. If these functions can be computed only once and then stored, we can avoid performing costly integrals or finite element calculations every time we have to solve Eqs.~\eqref{eqn:projections} for a new set of parameters $\alpha$. This property is exploited later when we construct an emulator for the Gross-Pitaevskii equation in Eq.~\eqref{eqn:G.P.}.

To illustrate the application of the RBM and connect with previous results in the EC literature, we work with the two previously mentioned examples for $F_\alpha$. For the single-channel scattering example, we create the approximate solution $\hat{\phi}_\alpha=\sum_{k=1}^na_k\phi_k$, with $\phi_k$ as exact solutions to $n$ different $\alpha$, and satisfy the boundary conditions by imposing $\sum_{k=1}^na_k=1$, as done in~\cite{furnstahl2020efficient}. By letting $a_1=1-\sum_{k=2}^na_k$ and thus $\hat{\phi}_\alpha=\phi_1+\sum_{k=2}^na_k(\phi_k-\phi_1)$, we explicitly satisfy the boundary condition while having only $n-1$ free coefficients. We identify $(\phi_k-\phi_1)$ for $k\geq 2$ as the relevant elements in the basis expansion and select $\psi_k=(\phi_k-\phi_1)$ as the associated projecting functions, leading to the equations:
\begin{equation}\label{eqn:affine galerkin}
  \sum_{k=1}^na_k\langle \phi_j-\phi_1|F_\alpha({\phi}_k)\rangle=0,\quad \text{for}\quad  2\leq j \leq n.
\end{equation}
These equations---formulated from a geometric projection argument---are equivalent to those obtained by the Kohn variational principle~\cite{lucchese1989anomalous}, as done in~\cite{furnstahl2020efficient,drischler2021toward}. The proof is elaborated in the supplementary material. 

In the bound system example, with $\psi_k=\phi_k$, Eq.~\eqref{eqn:projections} might not have a solution for the exact eigenvalue. Allowing $\lambda_\alpha$ to be approximated by $\hat{\lambda}_\alpha$ helps ensure we can solve the projected equations:
\begin{equation}
    \sum_{k=1}^ n a_k\langle\phi_j|H_\alpha|{\phi_k}\rangle=\hat{\lambda}_\alpha\sum_{k=1}^n a_k\langle \phi_j|{\phi}_k\rangle,\quad \text{for all }j,\label{eqn: projected hamiltonian}
\end{equation}
where the set $a_k$ plus the approximate eigenvalue $\hat\lambda_\alpha$ add up to $(n+1)$ unknowns. We can complete the set of equations with a normalization condition: $ \langle\hat{\phi}_\alpha|\hat{\phi}_\alpha\rangle=1$. When choosing $\phi_k$ as exact solutions for different $\alpha$, Eq.~\eqref{eqn: projected hamiltonian} is equivalent to the generalized eigensystem of EC formulated from the Rayleigh-Ritz variational method~\cite{frame2018eigenvector}.

Beyond these two examples, the generality of the Galerkin formalism allows to apply RBMs to a wide variety of problems, including discrete, operator, integral, and differential equations~\cite{fletcher1984computational,singh1974unified}. As such, the projected equations~\eqref{eqn:projections} can be directly applied to non-linear problems like non-linear eigenvalue equations where $F_\alpha = G_\alpha({\phi}_\alpha) - \lambda_{\alpha}\phi_\alpha$, with $G_\alpha$ a general operator. Additionally, in the case of coupled equations---common in many-body physics---the formalism is easily extended by expanding each coupled function $\hat\phi^{(i)}_\alpha$ independently and finding the coefficients $a_k^{(i)}$ by enforcing Eq.~\eqref{eqn:projections} with projecting functions $\{\psi_j^{(i)}\}_{k=1}^n$ for each $i$-th coupled function.

It is important to note, however, that if the solution manifold $\phi_\alpha$ for the problem at hand cannot be sufficiently embedded in a \textit{linear} subspace, then the RBM we described will not constitute an effective emulator. We can engineer a simple example with the 1D quantum Hamiltonian of a particle trapped in an infinite well~\cite{cohenV1} by letting $\alpha$ control the location of the well. A direct application of the RBM fails to accurately emulate the ground state wavefunction as $\alpha$ changes. Extensions to the basic methodology can tackle these issues by allowing further manipulation of the reduced basis~\cite{nonino2019overcoming}.

In practice, to test whether a problem is fit for emulation via RBMs, it is sufficient to observe the decay of the singular values $\sigma_k$ associated with the PCA of a group of exact solutions $\tilde\phi_\alpha$ for various $\alpha$~\cite{quarteroni2015reduced}. Fig.~\ref{fig:SVD} shows the singular values $\sigma_k$ for the problems discussed in our work, all decaying exponentially except for the infinite well. In the case of the scattering wavefunctions in the $^1S_0$ channel at a fixed energy for the Minnesota potential~\cite{thompson1977systematic}, the $\sigma_k$ are consistent with the EC results~\cite{furnstahl2020efficient}. A similar pattern is obtained for wavefunctions across energies (black squares in Fig.~\ref{fig:SVD}) by re-scaling the scattering differential equation via the change of variable $s=pr$, making all exact solutions $\tilde\phi_\alpha(s)$ share the same asymptotic behavior. This observation implies that in principle it should be possible to build a scattering emulator across energies. For the Gross-Pitaevskii equation and the 13 energy levels of $^{48}$Ca under density functional theory (DFT), the decay of their respective $\sigma_k$ also makes them excellent candidates for the application of the RBM, as we explore next.

The Gross-Pitaevskii equation~\cite{Gross:1961,Pitaevskii:1961} (see also~\cite{pichi2020reduced} for a RBM application) is a nonlinear Schrödinger equation that approximately describes the low-energy properties of dilute Bose-Einstein condensates. Using a self-consistent mean field approximation, the many-body wavefunction is reduced to a description in terms of a single complex-valued wavefunction $\phi(\vec{r})$. We work with the one-dimensional Gross-Pitaevskii equation~\cite{Carr:1999,Carr:2000,Carr:2001,Torres-Vega:2017} with a harmonic trapping potential by letting $F_\alpha$ be:
\begin{equation}
  F_{q,\kappa}(\phi)= -\phi''+\kappa x^2\phi+q|\phi|^2\phi-\lambda_{q,\kappa}\phi=0,\label{eqn:G.P.}
\end{equation}
where $\kappa$, $q$, and $\lambda_{q,\kappa}$ are proportional to the strength of the harmonic trapping, the self-coupling of the wavefunction, and the ground state energy, respectively. $\phi(x)$ is a single variable function that depends on $x$ and it is normalized to unity. Note that, since this equation depends linearly on $\kappa$ and $q$, the projection Eqs.~\eqref{eqn:projections} that involve integrals in $x$ can be evaluated and stored for faster computation, leading to expressions of the form in Eq.~\eqref{eqn:affine_decomposition}. For example, the term associated with the harmonic trapping reads: $\kappa \langle \psi_j| x^2 | \hat \phi  \rangle= \kappa \sum _{k=1}^na_k\langle \psi_j| x^2 |\phi_k  \rangle =\kappa \sum _{k=1}^n a_k\int \psi_j (x) x^2  \phi_k (x) dx $.  
\begin{figure*}[t]
    \centering
    \includegraphics[width=0.98\textwidth]{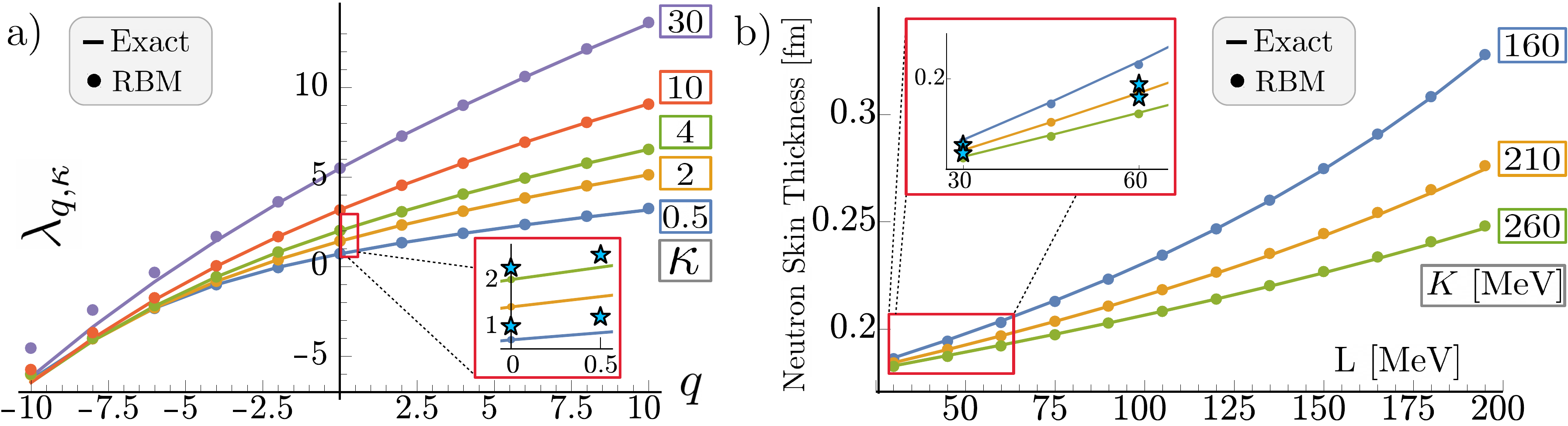}
    \caption{Comparison between the exact solvers (solid lines) and the RBM (points) calculations. Panel a) shows the ground state energy $\lambda_{q,\kappa}$ of the Gross-Pitaevskii equation as a function of $q$ for the values of $\kappa =[0.5,2,4,10,30]$. Panel b) shows the neutron skin thickness of $^{48}$Ca as a function of $L$ for the values of $K=[160,210,260]$ MeV. In both figures, the Lagrange basis with four points is shown as cyan stars within the inset plots. The supplementary material contains additional details on these calculations.}
    \label{fig:GSEnergy}
\end{figure*}

To test the RBM for extrapolation, we built a Lagrange basis with four training functions $\tilde\phi_i$ in the $[q,\kappa]$ space 
as exact solutions ($F_{q_i,\kappa_i}(\tilde\phi_i)=0$), with the projecting functions as $\psi_i=\phi_i$. Panel a) in Fig.~\ref{fig:GSEnergy} shows the results of emulating $\lambda_{q,\kappa}$ by using this basis and applying Eqs.~\eqref{eqn:projections} plus the normalization condition. The agreement between the exact and emulated calculations is excellent, with an error of less than $2.5\%$ in the repulsive phase ($q\geq 0$) where the four training parameters are located, and it deteriorates only in the attractive phase ($q<0$) well beyond the training region. Extrapolation is not a feature usually exploited on the RBM literature, yet it could be key when calculations of exact solutions in a specific phase of the system are numerically unstable or impossible, but approximable by such methods.

In addition to extrapolating, we explored a situation similar to how emulators are tested for UQ~\cite{konig2020eigenvector,drischler2021toward}. Using a Latin hypercube sampling (LHS)~\cite{mckay2000comparison}, we drew 500 testing points in the range $q\in[0,30]$ and $\kappa\in [5,30]$. We constructed three types of reduced basis: Lagrange, POD, and POD+Greedy, each with three sizes $n=(2,4,8)$. The Lagrange basis consisted of $n$ exact solutions drawn with LHS. The POD and POD+Greedy consisted of $n$ principal components from a set of $N=20$ exact solutions. For the POD the $N$ exact solutions were drawn using LHS, while for the POD+Greedy the first solution was placed at a central location and the other $N-1$ were included one-by-one through a Greedy algorithm inspired on Refs.~\cite{veroy2003posteriori,Haasdonk2008PODGreedy,Sarkar2021fpz}. Our Greedy approach finds the parameter set $[q_{m+1}, \kappa_{m+1}]$ for the next exact solution $\tilde\phi_{m+1}$, by maximizing the norm of the residual $F_{q,\kappa}(\hat \phi_{q,\kappa})$ over a LHS of parameters $[q,\kappa]$. In each step, $\hat\phi_{q,\kappa}$ is constructed with a POD basis informed by the previous $m$ exact solutions. 

Table~\ref{tab:results} shows the relative root mean squared errors (RMSEs), which converge exponentially as expected from the results in Fig.~\ref{fig:SVD}. Both POD bases were more accurate and robust than the Lagrange basis, which produced results that frequently changed by more than an order of magnitude when re-sampling the exact solutions for the basis. For $n=8$ the accuracy of the POD+Greedy basis was more than $600$ times better than the Lagrange basis. In terms of speed-up when calculating the 500 testing points, the three reduced bases with $n=2$ were almost $150$ times faster than the exact solver, while $n=4$ and $n=8$ obtained speed-ups of 40 and $5$ times, respectively.

\begin{table}
\caption{RMSEs for the Gross-Pitaevskii and DFT problems described in the text. The RMSE is defined as $\langle \big[(A_\text{RBM}-A_\text{exact})/A_\text{exact}\big]^2\rangle^{1/2}$, where $A$ is the quantity being computed, and $\langle \rangle$ denotes average. Three cases of the reduced basis size were explored with $n=(2,4,8)$. 500 testing points were drawn in their respective parameter space, but for DFT 32 points were excluded from the statistics since the exact solver reported convergence problems. The supplementary material contains additional details on these calculations.} \label{tab:results}

\begin{tabular}{c|ccc|c}
\hline
\multirow{2}{*}{\begin{tabular}[c]{@{}c@{}}Basis \\ n \end{tabular}} & \multicolumn{3}{c|}{\begin{tabular}[c]{@{}c@{}}Gross-Pitaevskii Ground\\ State Energy\end{tabular}}                                                                       & \begin{tabular}[c]{@{}c@{}}$^{48}$Ca Average\\ Particle Energy\end{tabular} \\
                                                                   & \multicolumn{1}{c|}{Lagrange} & \multicolumn{1}{c|}{POD} & \begin{tabular}[c]{@{}c@{}}POD\\ Greedy\end{tabular} & POD                                                                         \\ \hline
2                                                                  & $1.0\times 10^{-1}$                                                         & $1.2\times 10^{-2}$      & $1.5\times 10^{-2}$                                  & $5.9\times 10^{-3}$                                                                           \\
4                                                                  & $3.0\times 10^{-3}$                                                         & $5.6\times 10^{-4}$      & $2.1\times 10^{-4}$                                  & $6.1\times 10^{-4}$                                                                           \\
8                                                                  & $1.3\times 10^{-5}$                                                         & $1.2\times 10^{-6}$      & $2.0\times 10^{-8}$                                  & $1.7\times 10^{-4}$                                                                           \\ \hline
\end{tabular}
\end{table}

We now proceed to use the RBM in realistic nuclear DFT calculations. DFT is a widely applied microscopic formalism~\cite{nucleardft} (see also~\cite{cances2007feasibility,lin2012adaptive,zhang2017adaptive} for other RBM applications to DFT). In nuclear physics it is used to describe properties of nuclei from the mean-field perspective, i.e., each nucleon interacts with an average effective field made up of all the particles in the system. This interaction is then constructed in a self-consistent way: the wavefunction of each nucleon and its eigenenergy are found at the same time as the effective field they produce and interact with. As such, the Hamiltonian $\hat{h}^{(i)}$ acting on the $i$-th wavefunction $\phi^{(i)}$ depends on all $M$ of them: 
\begin{equation}\label{eq: Hamiltonian DFT}
    \hat{h}^{(i)}[\Phi]\phi^{(i)} - \lambda^{(i)} \phi^{(i)}=0 \quad \text{for}\quad  1\leq i \leq M,
\end{equation}
where $\Phi = \{\phi^{(i)}\}_{i=1}^M$, and the parameter list $\alpha$ has been omitted for the sake of clarity. The dependence of the Hamiltonian on the wavefunctions comes from, for example, the total nuclear density $\rho$ and kinetic energy density $\tau$. We derive the single particle Hamiltonian, $\hat{h}^{(i)}$, from the Skyrme effective interaction~\cite{skyrme,skyrme1958,vautherin1972,engel1975},
the nuclear part of which can be written as a general energy density functional (EDF) of time-even densities~\cite{dobaczewski1995}:
\begin{equation}
\label{eq:edf}
    {\cal H}_t(r)= C_t^{\rho}\rho_t^2
    + C_t^{\rho\Delta\rho}\rho_t\Delta\rho_t
    + C_t^{\tau}\rho_t\tau_t
    + C_t^{   J} \tensor{J}_t^2
    + C_t^{\rho\nabla J}\rho_t\nabla\cdot\mathbf{J}_t,
\end{equation}
where the subscript $t=(0,1)$ represents isoscalar and isovector densities, respectively.
The parameters of this EDF, $C_t^{\tau}$ for instance, model the coupling between the particles and the nucleonic density in question (the kinetic energy density, $\tau$, in this case). As it is usually done in modern EDF optimization~\cite{chabanat1997,kortelainen2010}, we can parametrize those couplings in terms of nuclear-matter properties plus the remaining coupling constants left unconstrained:
\begin{multline}
\left\{
\rhoc, \enm/A, \knm, \asym, \lsym, M_s^{*},
C_{t}^{\rho\Delta\rho}, C_{t}^{\rho\nabla J}
\right\}.
\label{parameter_set}
\end{multline}
This representation is primarily rooted in physical observables---like the nuclear saturation density, $\rho_c$---and simplifies the selection of a sensible range of values to explore in model calibration for DFT, and for constructing the training bases for the RBM.

To test the RBM in extrapolation for DFT, we built a Lagrange basis of four points spanning $\lsym=[30,60]$ MeV and $\knm=[200,220]$ MeV while the other parameters remained at their optimized UNEDF1 values~\cite{kortelainen2011}. The wavefunctions on each shell (7 for neutrons and 6 for protons) were calculated using both the exact solver and the RBM emulator. Panel b) in Fig.~\ref{fig:GSEnergy} shows the performance of the emulator when calculating the neutron skin thickness of $^{48}$Ca~\cite{hagen2016neutron}, a quantity particularly sensitive to the $\lsym$ parameter.
The agreement between the emulated values and exact DFT results is excellent, with an error of less than $0.8\%$ for all extrapolated values shown, even for $\lsym$ and $\knm$ well outside the training zone.

To test the limits of the emulator, the range of all ten available parameters in Eq.~\eqref{parameter_set} were widened well beyond what is reasonable for realistic nuclear matter. We used LHS to draw 50 training points to build a POD basis with $n=(2,4,8)$ and to independently draw 500 testing points within the widened parameter ranges. As such, several parameter combinations yielded convergence issues for the DFT solver, but not for the emulated calculations, highlighting the capability of RBMs to extrapolate into regions where exact solvers can experience numerical instabilities. Even though the emulated results of non-converging test points seemed reasonable, we consider their validation to be beyond the scope of this work.

As Table~\ref{tab:results} shows, for the stable parameter sets, the RBM reproduces single nucleon energies well. This is particularly striking for the reduced basis with only two elements, which gives an error of about $0.6\%$ despite all ten parameters being varied in the test sample. In terms of speed-up when calculating the 500 testing points, the reduced basis with $n=(2,4,8)$ were $6$, $4$, and $2$ times faster than the exact solver, respectively. We note that these speedups were obtained without precomputing any of the terms involved in Eq.~\eqref{eqn:projections}. Greater speed-ups can be achieved by precomputing as many of the terms in Eq.~\eqref{eqn:projections} as possible, in the traditional strategy of an offline/online procedure often seen in RBM applications. Indeed, by separating the Hamiltonian in Eq.~\eqref{eq: Hamiltonian DFT} into the parts that can and cannot be precalculated (called affine and non-affine in the RBM literature~\cite{quarteroni2015reduced}), we achieve speed-ups of more than 250 times with respect to the exact solver for a reduced basis of two elements. 

The parts of the Hamiltonian~\eqref{eq: Hamiltonian DFT} that are non-affine in the parameters can be made affine by using techniques such as the Empirical Interpolation Method~\cite{barrault2004empirical,grepl2007efficient}. The terms that are nonlinear in the wavefunctions on the other hand, such as powers of the density $\rho$, can present a problem due to a combinatorially increasing terms ($M_k$) in Eq.~\eqref{eqn:affine_decomposition}. We will further study these challenges in a future work.

Speed-up gains of more than two orders of magnitude will enable large scale model UQ studies for a wide range of EDFs \cite{giuliani2022bayes}, an endeavor which, up to now, seemed inaccessible. Furthermore, the RBM approach could also reduce the penalty of using higher-dimensional solvers for systematic studies and UQ, calculations previously limited to spherical and cylindrical symmetries. Finally, the trained emulators could be deployed in a cloud computing environment \cite{bmex}, fostering collaborative research and facilitating the expansion of the scientific network. 

We hope our results help spark the interest of the nuclear theory community in RBMs. For this purpose, we created and will continue to update an online resource~\cite{rbmNuclear} to illustrate many of the concepts we discussed. The adoption of recent developments on the choice of ansatz subspaces~\cite{benner2015survey,OHLBERGER2013901,nonino2019overcoming}, on error bounds and convergence properties~\cite{veroy2003posteriori,buffa2012priori,cohen2020greedy}, and on the computational efficiency for non-affine and nonlinear problems~\cite{barrault2004empirical,grepl2007efficient}, to name a few, could become key in reaching the full extent of what these methods can offer. Given the simplicity and flexibility of the Galerkin projection, and the PCA diagnostic we showcase to test for low-dimensional manifolds, we believe that RBMs have the potential to become standard tools for the emulation of challenging problems in many-body nuclear physics. 

\begin{acknowledgments} 
\hypersetup{bookmarksdepth=-1}
\section*{Acknowledgements}

We are very grateful to Witek Nazarewicz and Frederi Viens for their critical observations during the elaboration of this manuscript. We also thank Ana Posada, Jorge Piekarewicz, and Daniel Phillips for their careful read of the manuscript. This work was supported by the National Science Foundation CSSI program under award number 2004601 (BAND collaboration), the U.S. Department of Energy under Award Number DOE-DE-NA0003885 (NNSA, the Stewardship Science Academic Alliances program), U.S. Department of Energy (DE-SC0013365 and DE-SC0021152) and the Nuclear Computational Low-Energy Initiative (NUCLEI) SciDAC-4 project (DE-SC0018083).
\end{acknowledgments}

\bibliographystyle{apsrev4-1}
\bibliography{gevc} 

\clearpage

\appendix
\setcounter{table}{0}
        \renewcommand{\thetable}{S\arabic{table}}%
        \setcounter{figure}{0}
        \renewcommand{\thefigure}{S\arabic{figure}}%
        \setcounter{equation}{0}
        \renewcommand{\theequation}{S\arabic{equation}}%
        \setcounter{algocf}{0}
        \renewcommand{\thealgocf}{S\arabic{algocf}}
        
\section{Supplementary Material}

\section{Equivalence of the Reduced basis method and Eigenvector Continuation under the Kohn Variational Principle}

For the sake of conciseness, we only treat the case of eigenvector continuation applied to single-channel scattering, as done in \cite{furnstahl2020efficient} using the Kohn Variational principle. The extension to the generalized Kohn principle \cite{drischler2021toward,lucchese1989anomalous} is straightforward and follows the same steps shown below.

Within this context, we aim at finding approximate solutions to a differential equation of the form \cite{thompson2009nuclear}:
\begin{equation}
    F_\alpha(\phi)=\left(-\frac{d^2}{dr^2}+\frac{\ell(\ell+1)}{r^2}+U(r,\alpha)-p^2\right)\phi(r)=0,\label{eqn:scattering_eqn}
\end{equation}
where $U(r,\alpha)$ represent the potential the particle interacts with, which we assume changes smoothly with its parameter set $\alpha$, $\ell$ is the angular momentum quantum number and $p$ relates to the asymptotic momentum. Let us assume that the solution to this equation is subject to the boundary conditions $\phi(r=0)=0$, and
 \begin{equation}
    \phi(r)\underset{r\rightarrow\infty}{\longrightarrow} \frac{1}{p}\sin\left(pr-\ell\frac{\pi}{2}\right)+\tau\cos\left(pr-\ell\frac{\pi}{2}\right).\label{eqn:asymptotic_behavior}
\end{equation}
Note that Eq.~\eqref{eqn:asymptotic_behavior} imposes a normalization condition on $\phi$: the coefficient accompanying the sine function must equal $1/p$. 

A straightforward application of the reduced basis method (RBM) as discussed in the main document leads to the choice of an approximate function:
\begin{equation}
    \hat{\phi}_\alpha=\sum_{k=1}^n a_k\phi_k,\quad\text{with}\quad \sum_{k=1}^n{a_k}=1,\label{eqn:galerkin_scattering_1}
\end{equation}
where the $\phi_k$ are solutions to $F_{\alpha_k}\phi=0$ with the correct boundary conditions. Following the discussion on the main text, we eliminate the redundancy of the coefficients $a_k$ created by the boundary conditions by explicitly writing one of the them in terms of the others. Without loss of generality, we let $a_1=1-\sum_{k=2}^na_k$, obtaining:
\begin{equation}
    \hat{\phi}_\alpha=\phi_1+\sum_{k=2}^n a_k(\phi_k-\phi_1).\label{eqn:galerkin_scattering_2}
\end{equation}

With this rewriting, it becomes explicit that we only need to determine $(n-1)$ coefficients to obtain the approximate solution, which is part of the affine space spanned by $(\phi_k-\phi_1)$. Following a similar reasoning to the main text, we choose $\psi_k=({\phi}_k-\phi_1)$ for $k=2,\dots,n$, resulting in the following equations for the remaining $a_k$:
\begin{align}
\begin{split}
& \langle\psi_j|F_\alpha(\hat\phi_\alpha)\rangle=  \langle{\phi}_j-\phi_1|F_\alpha(\phi_1)\rangle \ +\\
&\sum_{k=2}^na_k\langle{\phi}_j-\phi_1|F_\alpha({\phi}_k-\phi_1)\rangle=0, \label{eqn:galerkin_scattering_intermediate}
\end{split}
\end{align}
for $j=2\dots,n,$ where the inner product is defined as $\langle\psi|\phi\rangle=\int_0^\infty\psi(r)\phi(r)dr$ (without the usual complex conjugation).

We can rewrite Eq.~\eqref{eqn:galerkin_scattering_intermediate} in a more compact form by reintroducing $a_1$, and the normalization condition. This results in the equation set:
\begin{flalign}
\sum_{k=1}^na_k\langle{\phi}_j-\phi_1|F_\alpha(\phi_k)\rangle=0,\quad\text{for}\quad j=2,\dots,n;\label{eqn:galerkin_scattering_final}\\
\text{and}\quad a_1=1-\sum_{k=2}^na_k.
\end{flalign}

\subsection{Approximation through the Kohn Variational Principle}
The ($K$-matrix) Kohn variational principle (KVP) states that the solution to Eq.~\eqref{eqn:scattering_eqn} with the asymptotic behavior \eqref{eqn:asymptotic_behavior}, is a stationary point for the functional:
\begin{equation}
    \beta[\phi]=\tau[\phi]-\int_0^\infty dr\phi(r) F_\alpha\left(\phi(r)\right),\label{eqn:KVP_1}
\end{equation}
where $\tau[\phi]$ extracts its value from the asymptotic behavior of $\phi$, that is, the cosine coefficient in Eq.~\eqref{eqn:asymptotic_behavior}.

The Kohn variational method is detailed in the supplemental material of Ref. \cite{furnstahl2020efficient}. It utilizes a trial function $\hat\phi_\alpha$ constructed exactly as in Eq.~\eqref{eqn:galerkin_scattering_1}, and it finds a stationary point of the functional \eqref{eqn:KVP_1} in terms of the coefficients $a_k$. After using the method of Lagrange multipliers to enforce the normalization condition, it is found that the following equation set describes the stationary point:
\begin{equation}
    \tau_j-\lambda-\sum_{k=1}^n(\Delta U_{jk}+\Delta U_{kj})a_k=0\quad \text{for}\quad j=1\dots,n;\label{eqn:KVP_lagrange}
\end{equation}
where $\tau_j$ is the cosine coefficient in Eq.~\eqref{eqn:asymptotic_behavior} associated with each $\phi_j$, and the matrix $\Delta U_{jk}$ is a shorthand notation for the inner products:
\begin{equation}
    \Delta U_{jk}=\langle\phi_j|F_\alpha(\phi_k)\rangle.\label{eqn:delta_U}
\end{equation}
These equations, together with the normalization condition $\sum_{k=1}^n{a_k}=1$, can be used to find the $n$ coefficients $a_k$, plus the Lagrange multiplier $\lambda$.

\subsection{Proof of equivalence between the methods}
To compare Eq.~\eqref{eqn:KVP_lagrange} to Eq.~\eqref{eqn:galerkin_scattering_final}, we need to rewrite Eq.~\eqref{eqn:KVP_lagrange} in terms of $\Delta U_{jk}$ only, eliminating both the $\tau_j$ and the Lagrange multiplier $\lambda$. We can relate the elements of $\Delta U_{kj}$ to their transposes $\Delta U_{jk}$ by integration by parts. Note that
\begin{equation}
    -\int_0^\infty\phi_k\phi''_j dr=-\left.\left(\phi_k\phi_j'-\phi_k'\phi_j\right)\right|_0^\infty-\int_0^\infty\phi_k''\phi_jdr,\label{eqn:by_parts}
\end{equation}
where the boundary term can be evaluated through the boundary conditions \eqref{eqn:asymptotic_behavior} to be $\tau_j-\tau_k$. Therefore, via integration by parts, we can make $F_\alpha$ act on $\phi_j$ in Eq.~\eqref{eqn:delta_U} to obtain:
\begin{equation}
\Delta U_{kj}=\Delta U_{jk}+(\tau_j-\tau_k).\label{eqn:delta_U_transpose}
\end{equation}

Using this result we can eliminate $\Delta U_{kj}$ from Eq.~\eqref{eqn:KVP_lagrange} and obtain:
\begin{align}
\begin{split}
    \tau_j-\lambda-\sum_{k=1}^n(2\Delta U_{jk})a_k- \sum_{k=1}^na_k(\tau_j-\tau_k)=0\\ 
    -\lambda-\sum_{k=1}^n(2\Delta U_{jk})a_k+ \sum_{k=1}^na_k\tau_k=0,
    \label{eqn:KVP_lagrange_2}
\end{split}
\end{align}
for $j=1,\dots,n$; where we used the fact that the $a_k$ sum to unity to cancel the $\tau_j$ term. 

Next, we can eliminate the Lagrange multiplier $\lambda$ and the sum of $a_k\tau_k$ by subtracting equations for different $j$. Without loss of generality, we can subtract all equations to the equation corresponding to $j=1$, resulting in:
\begin{equation}
    \sum_{k=1}^n(\Delta U_{jk}-\Delta U_{1k})a_k=0\quad\text{for}\quad j=2\dots,n.\label{eqn:KVP_lagrange_3}
\end{equation}
Finally, using the definition of $\Delta U_{jk}=\langle\phi_j|F_\alpha(\phi_k)\rangle$ shows the equivalence with Eq.~\eqref{eqn:galerkin_scattering_final}, completing the proof.

It is interesting to note that the second term in Eq.~\eqref{eqn:asymptotic_behavior} can be thought as a first-order correction to the emulated $\tau$. This correction adds a factor proportional to the residual $F_\alpha(\hat{\phi}_\alpha)$, in a similar spirit as Newton's method \cite{ascher1995numerical}. The `proportionally factor' (or transfer function \cite{giuliani2021noise}) connecting the residual $|F_\alpha(\hat\phi) \rangle$ and $\tau$ is in this case the negative of the exact solution $-\langle\phi|$. This in turn, can be approximated by the emulated solution $-\langle\hat\phi|$, obtaining the correction:
\begin{equation}
    -\langle\hat\phi|F_\alpha(\hat\phi)\rangle= -\int_0^\infty dr\phi(r) F_\alpha\left(\phi(r)\right).
\end{equation}

Such first order correction, together with the POD+Greedy algorithm we used in the main text, highlight the fact that the residual $F_\alpha(\hat\phi)$ is information rich and can be useful to improve the performance of the RBM emulator. 
\section{Reduced Basis Method on a set of coupled equations}

As we described in the main text, the RBM can be applied to the case with a set of $m$ coupled equations:
\begin{equation}
    F_\alpha^{(i)}(\phi^{(1)}_\alpha,\dots,\phi^{(m)}_\alpha)=0,\quad\text{for}\quad i=1,\dots,m;\label{eqn:nonlinear_coupled}
\end{equation}
by approximating each $\phi_\alpha^{(i)}$ as a linear combination of their corresponding solutions for different values of $\alpha$:
\begin{align}
\begin{split}
    & \hat{\phi}^{(i)}_\alpha=\phi_0^{(i)}+\sum_{k=1}^na^{(i)}_k\phi_k^{(i)},\\
    &\text{for}\quad k=1,\dots,n;\quad i=1,\dots,m.\label{eqn:training_f_coupled}
\end{split}
\end{align}

By selecting $\psi_j^{(i)}$ as the generators of the affine spaces for each $\hat\phi^{(i)}_\alpha$, we obtain the $n\times m$ Galerkin equations:
\begin{equation}
    \langle \psi_j^{(i)}|F^{(i)}_\alpha(\phi^{(1)}_\alpha,\dots,\phi^{(m)}_\alpha)\rangle=0, \label{eqn:projections_coupled}
\end{equation}
for $j=1,\dots,n;\ \text{and}\ i=1,\dots,m;$ to obtain the coefficients $a_k^{(i)}$. 

In the case of coupled eigenvalue-eigenvector systems, we can proceed as in the case of a single equation, by substituting the $m$ eigenvalues $\lambda^{(i)}_\alpha$ for approximate values $\hat\lambda^{(i)}_\alpha$, and enforcing $m$ normalization conditions, in accordance with the requirements of the problem at hand.

\section{Details about the numerical results}

The codes used to generate all the results we presented in the main text and here can be found in \cite{rbmNuclearCodes}.
\subsection{Decay of singular values}

\begin{figure}[hb]
    \centering
    \includegraphics[width=0.45\textwidth]{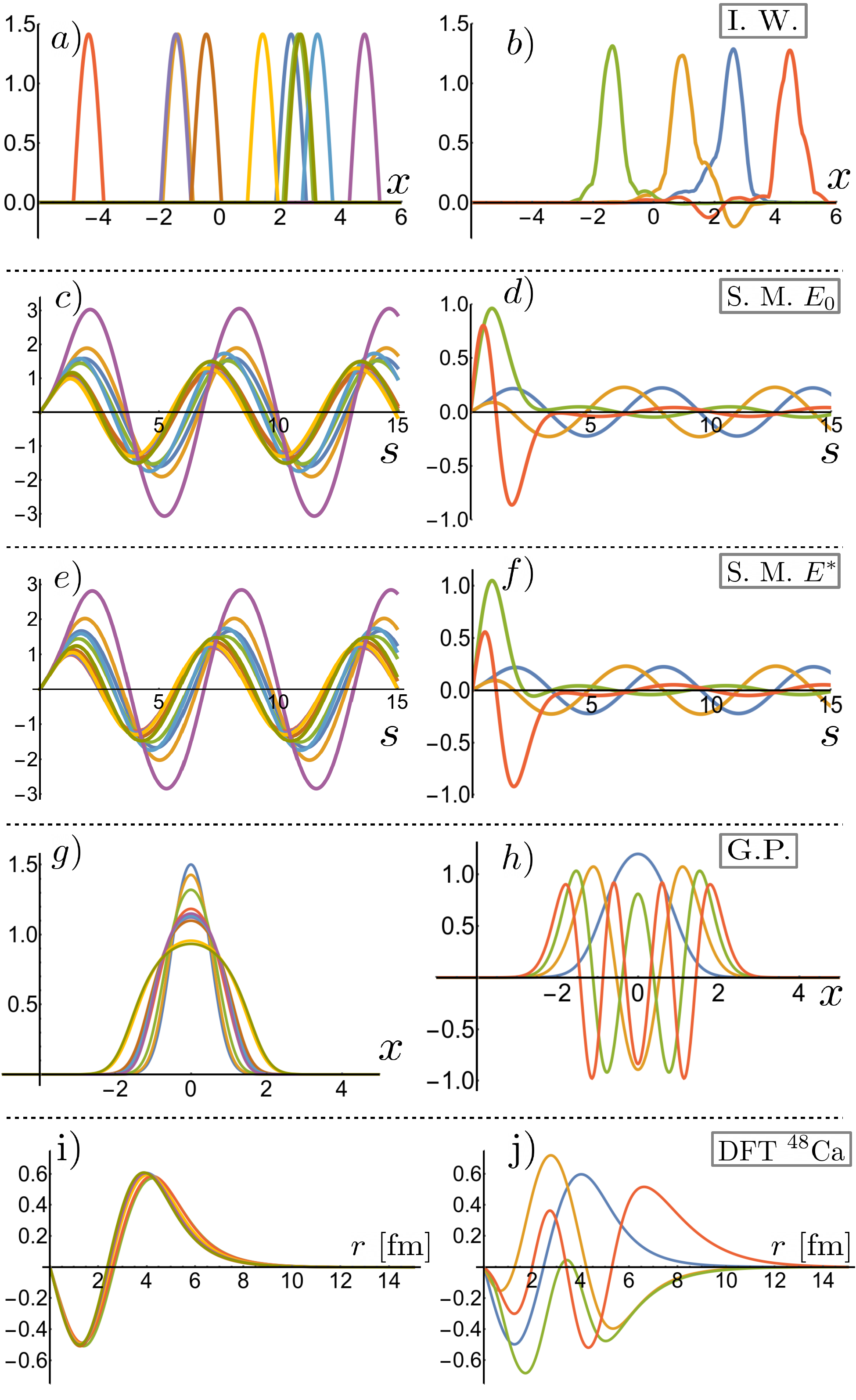}
    \caption{Examples of solutions (left column) and principal components (right column) for all the problems discussed in Fig.~1 in the main text: panels a) and b) correspond to the Infinite Well; panels c) and d) to the Scattering with the Minnesota potential at a fixed energy; panels e) and f) to the Scattering with the Minnesota potential with a variable energy; panels g) and h) to the Gross-Pitaevskii equation; and panels i) and j) to $^{48}$Ca under DFT. For panels i) and j) we chose the $^2S_{1/2}$ neutron wave function as a showcase, out of the total of 13 distinct levels for protons and neutrons. The left column shows exact solutions for 10 different values of the respective sampled parameters, while the right column shows the first 4 principal components obtained from a SVD analysis on a set of 40 exact solutions. }
    \label{fig: 4 examples}
\end{figure}

In this section, we give details on the construction of the singular values $\sigma_k$ from the singular value decomposition (SVD) showed in Fig.~1 of the main text. Four problems were considered: the infinite well (IW), the single channel 2-body scattering with a Minnesota potential at fixed energy (SM$E$) and varying energy (SM$E^{*}$), the Gross-Pitaevskii equation (GP), and $^{48}$Ca under density functional theory (DFT). Fig~\ref{fig: 4 examples} shows, for all problems considered, solutions for 10 different values of their parameters (left column), and the first 4 principal components out of a sample of 40 parameters in each case (right column). 

The IW problem consists of the 1D quantum Hamiltonian of a particle trapped in an infinite well (IW) \cite{cohenV1} where $\alpha$ controls the location of the well:
\begin{equation}
    V(x,\alpha) = \begin{cases}
0, & \alpha < x <\alpha+1,\\
\infty, & \text{otherwise.}
\end{cases}
\end{equation}
The ground state solutions to this Hamiltonian are wave functions of the form $\phi_\alpha=\sqrt{2}\sin{[\pi (x-\alpha)]}$ for $ \alpha < x < \alpha +1$ and zero otherwise. The singular values $\sigma_k$ for the IW showed in Fig.~1 of the main text were obtained by sampling 40 values of $\alpha$ in the range $[-5,5]$ using Latin hypercube sampling (LHS), and performing SVD on the set of 40 solutions. The $\sigma_k$ do not decay exponentially, and as can be seen in Fig.~\ref{fig: 4 examples} the first 4 principal components are unable to capture the variability of the set of solutions. As mentioned in the main text, extensions to the basic RBM  \cite{nonino2019overcoming}, such as allowing the training functions to be re-scaled and shifted in their $x$ domain, are able to tackle these issues. In this particular example, taking advantage of the symmetry of the problem, i.e.,  $\phi_\alpha(x+\alpha)=\phi_{\alpha=0}(x)$, would lead to all $\sigma_k$ to be zero for $k\geq 2$.

Both SM$E$ and SM$E^{*}$ problems consist of the single-channel $^1S_0$ nucleon-nucleon scattering Hamiltonian \cite{thompson2009nuclear}. The Minnesota potential \cite{thompson1977systematic} with parameters $[V_{0R},V_{0S}]$ is used for the interaction:
\begin{equation}
    V(r,V_{0R},V_{0S})=V_{0R}e^{-1.487r^2}+ V_{0S}e^{-0.465r^2},
\end{equation}
where the other two non-linear parameters were fixed. In both cases, we make the change of variables $s=pr$ and the scattering Hamiltonian takes the form:
\begin{equation}
    \left(-\frac{d^2}{ds^2}+\frac{\ell(\ell+1)}{s^2}+\tilde U(s,\alpha,p)-1\right)\phi_\alpha(s)=0,
\end{equation}
where the potential $U(r,\alpha)=V(r,\alpha)2\mu$ is now momentum dependent: $\tilde U(s,\alpha,p)=U(s/p,\alpha)/p^2$. In both cases, (SM$E$ and SM$E^{*}$) 40 parameters where obtained by a LHS in the range $V_{0R}=[100, 300]$ MeV and $V_{0S}=[-200, 0]$ MeV, following \cite{furnstahl2020efficient}. The singular values showed in Fig.~1 of the main text were obtained by performing SVD on the set of 40 solutions. In the case of SM$E$, all 40 solutions shared the same energy in the center of mass $E=50$ MeV, while for SM$E^{*}$ the energies where equispaced in the range $E=[20,80]$ MeV. 

The GP and DFT cases are explained in detail in the main text. The ranges for the parameters in both cases correspond to the ones used in Table I of the main text. For GP, a set of 40 values of the parameters $[q,\kappa]$ were obtained by LHS in the range $q\in[0,30]$ and $\kappa\in [5,30]$. For DFT, 50 values of the parameters were obtained with a LHS across the parameter ranges shown in Table~\ref{tab:results SM}.

\subsection{1-D Gross-Pitaevskii equation with a harmonic trapping potential}

\begin{figure}
    \centering
    \includegraphics[width=0.45\textwidth]{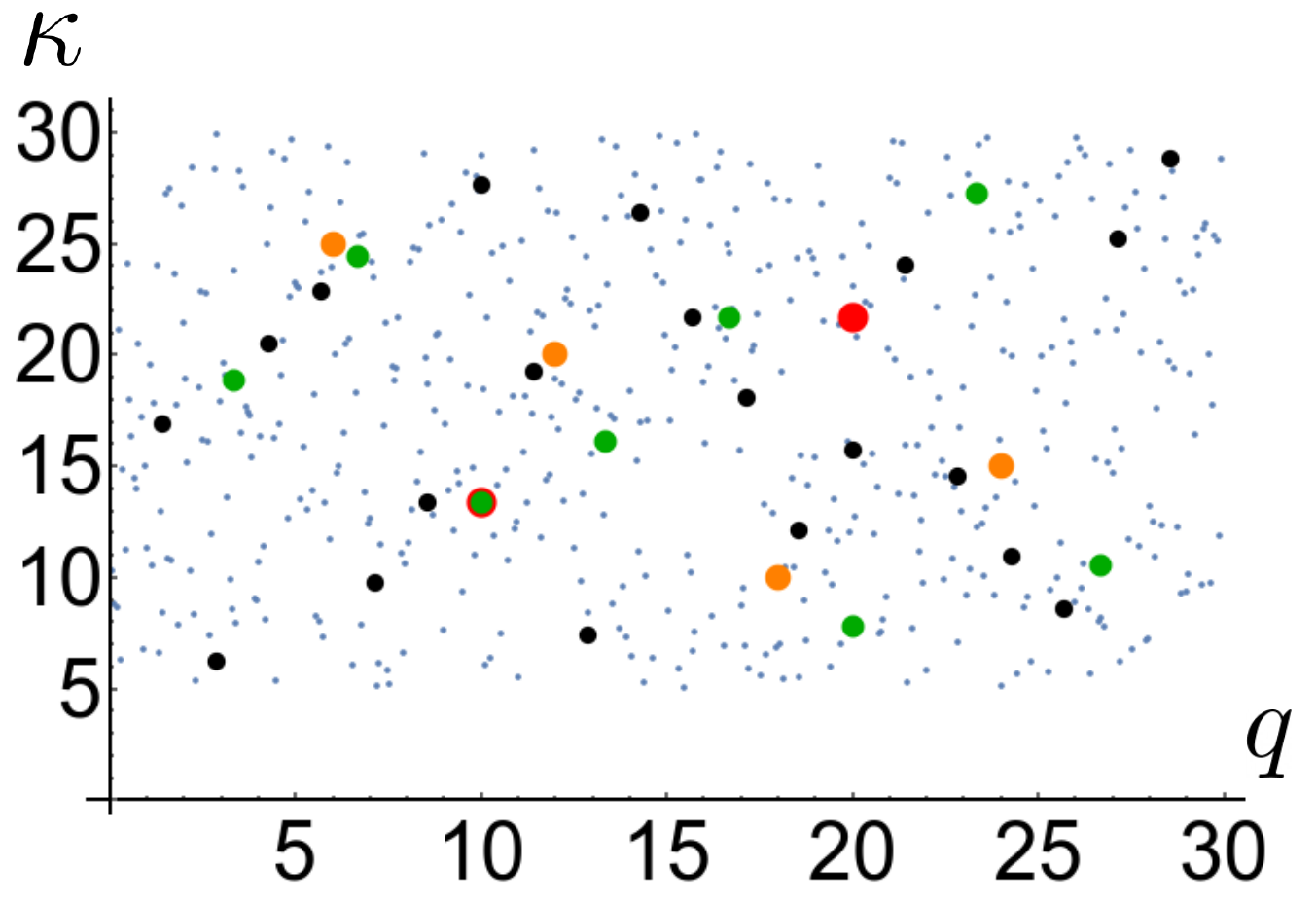}
    \caption{Parameter values used for the results of the GP equation in Table I of the main text, all drawn by Latin Hypercube Sampling \cite{mckay2000comparison}. Points in blue show the 500 testing samples. Points in red, orange, and green show the parameters for the Lagrange RB built with 2, 4, and 8 exact calculations, respectively. Points in black show the 20 parameters used to build the three POD RBs with $n=(2,4,8)$.}
    \label{fig:training points GP}
\end{figure}

\begin{figure*}[h]
    \centering
    \includegraphics[width=0.8\textwidth]{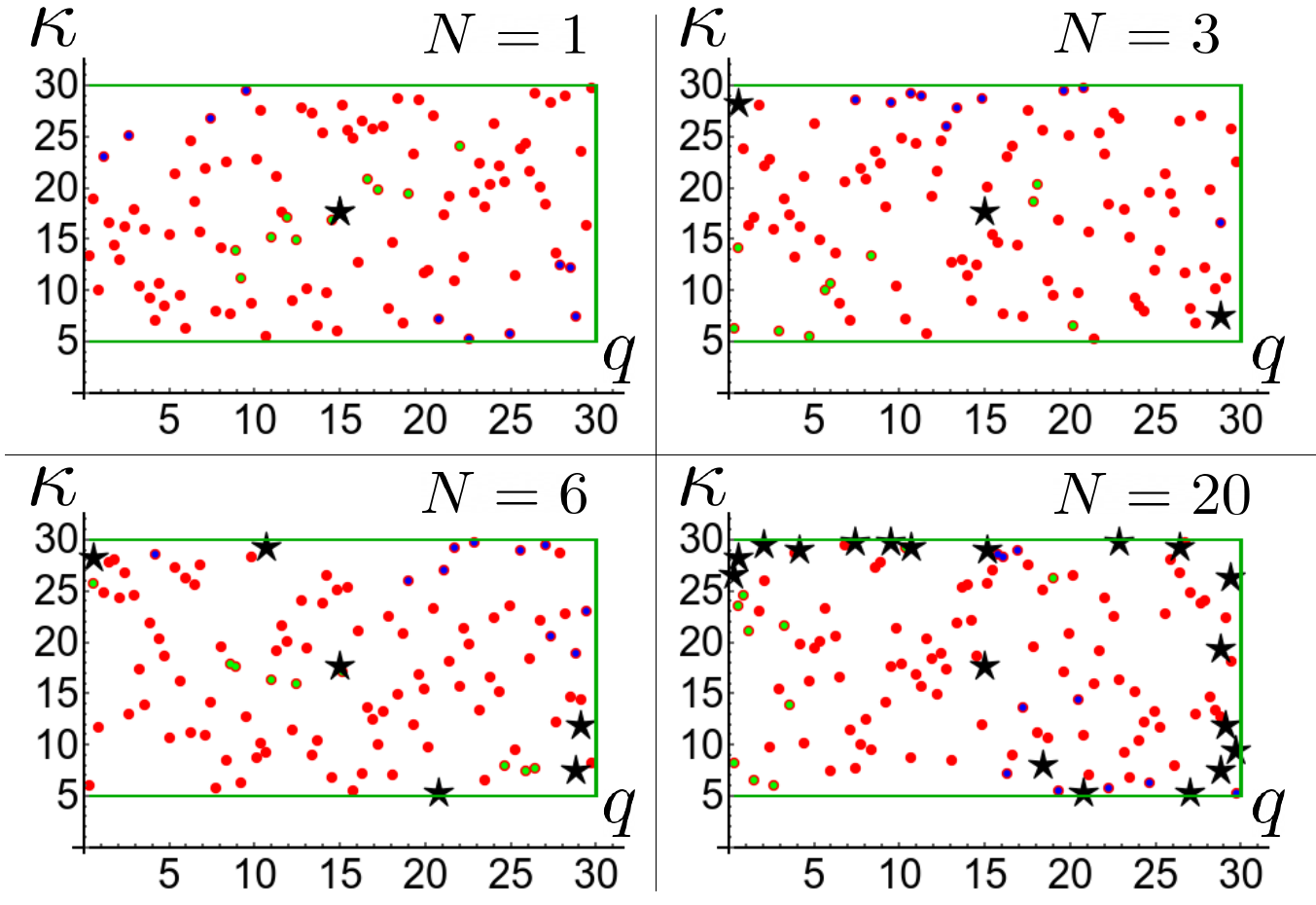}
    \caption{Construction of the POD+Greedy RB for the results in Table I of the main text. The black stars show the $N$ exact calculations made at each stage to build the POD+Greedy basis. The four panels show the stages for $N=(1,3,6,20)$. The red points in each panel are the 100 LHS draws on which the norm of the residual is maximized each time to add a new exact calculation to the POD+Greedy RB. This random sampling is repeated on each step (no two panels share the same red points). Points in blue and green on each panel show the 10 locations where the residual is maximized or minimized, representing the regions where the emulator is performing poorly and adequately, respectively. The green box shows the limits for the LHS: $q\in[0,30]$ and $\kappa\in [5,30]$.}
    \label{fig:training points GP 2}
\end{figure*}

\begin{figure*}[ht]
    \centering
    \includegraphics[width=0.9\textwidth]{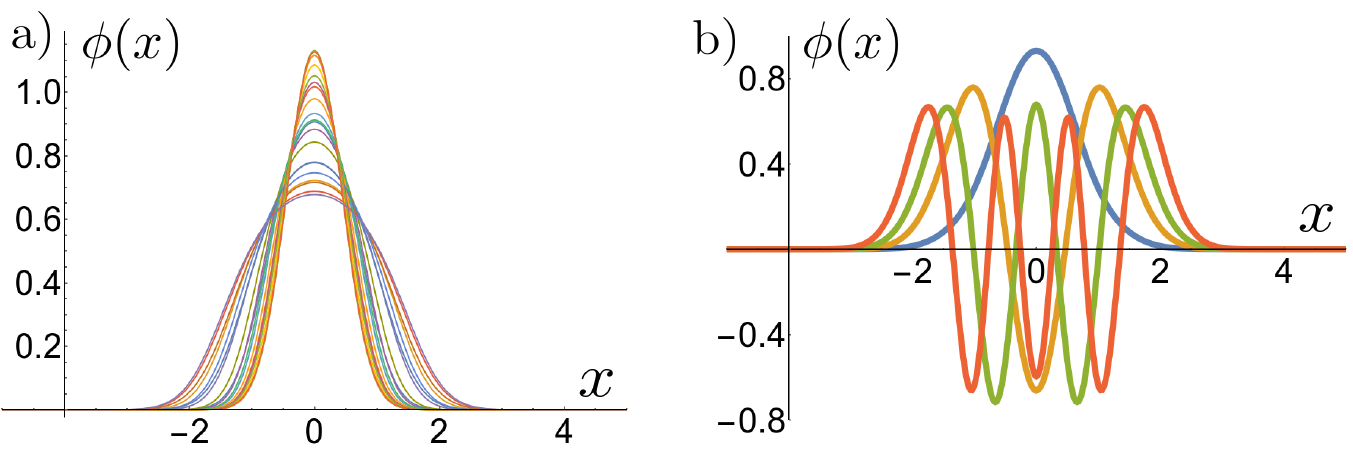}
    \caption{Panel a) shows the 20 exact solutions $\phi_{q,\kappa}(x)$ obtained by the Greedy algorithm in Fig.~\ref{fig:training points GP 2} and used to construct the POD+Greedy basis for the results of Table I in the main text. Panel b) shows the first four principal components of this set which constitute the RB used for the POD+Greedy results with $n=4$ in Table I in the main text.}
    \label{fig:SVD basis}
\end{figure*}

The four training points $\phi_k$ used in the Lagrange basis for the results of Fig.~2 a) in the main text are : $[q,\kappa]=\{[0, 1], [0, 5], [0.5, 1], [0.5, 5]\}$. Fig.~\ref{fig:training points GP} shows the training points for the Lagrange and POD RB, as well as the 500 testing points used for the results shown in Table I in the main text. Fig.~\ref{fig:training points GP 2} shows the construction on the POD+Greedy basis also used for the results shown in Table I in the main text. Fig.~\ref{fig:SVD basis} shows the 20 exact solutions $\phi_{q,\kappa}(x)$ selected by the Greedy algorithm, as well as the first four principal components of this set.
\begin{algorithm}
\caption{POD+Greedy scheme}\label{alg:greed}
Define starting parameters $\alpha_1$, $n_2$, $N$, $N_2$\\
Find $\tilde\phi_1$ s.t. $F_{\alpha_1}(\tilde\phi_1)=0$\\
$ES\gets\{\tilde \phi_1\}$\\
\For{$i=2,\dots,N$}{
$RB\gets POD(ES,n_2)$\\
Draw $N_2$ parameters with LHS: $A=\{\hat\alpha_1,\dots,\hat\alpha_{N_2}\}$\\
Use the RBM with the RB to find $\hat\phi_{\hat\alpha}$ for each element in  $A$\\
$\alpha_i\gets\argmax_{\hat\alpha\in A}||F_{\hat\alpha}(\hat\phi_{\hat\alpha})||^2$\\
Find $\tilde\phi_i$ s.t. $F_{\alpha_i}(\tilde\phi_i)=0$\\
$ES$ $\gets ES\cup\{\tilde \phi_i\}$}
\Return $ES$
\end{algorithm}
\begin{table}
\caption{Ranges for the 10 parameters used to generate the DFT results in Table I in the main text.} \label{tab:results SM}
\begin{tabular}{llll}
                                              & Min   & Max   & Units      \\ \hline
\multicolumn{1}{l|}{$\rho_c$}                 & 0.14 & 0.18 & fm$^{-3}$  \\
\multicolumn{1}{l|}{E$^\text{NM}$/A}          & -16.5 & -14.5 & MeV        \\
\multicolumn{1}{l|}{K$^\text{NM}$}            & 160   & 260   & MeV        \\
\multicolumn{1}{l|}{a$^\text{NM}_\text{sym}$} & 26    & 32    & MeV        \\
\multicolumn{1}{l|}{L$^\text{NM}_\text{sym}$} & 20    & 180    & MeV        \\
\multicolumn{1}{l|}{M$^{*}_s$}                & 0.7  & 1.4  &            \\
\multicolumn{1}{l|}{$C_0^{\rho \Delta \rho}$} & -55   & -40   & MeV fm$^5$ \\
\multicolumn{1}{l|}{$C_1^{\rho \Delta \rho}$} & -165  & -90   & MeV fm$^5$ \\
\multicolumn{1}{l|}{$C_0^{\rho \nabla J}$}    & -105   & -55   & MeV fm$^5$ \\
\multicolumn{1}{l|}{$C_1^{\rho \nabla J}$}    & -50   & -15     & MeV fm$^5$
\end{tabular}
\end{table}

\subsubsection{Description of the POD+Greedy scheme used for GP}
The POD+Greedy scheme used for the results of Table I of the main text consists on iteratively constructing a set of $N$ exact solutions ($ES$) with a (weak) Greedy algorithm \cite{quarteroni2015reduced} informed by the residuals of a POD RB of dimension $n_2$ derived from the set of exact solutions at each step. To set up the algorithm, let $POD(\{\tilde\phi_l\}_{l=1}^N,n)$ be a function that returns a normalized basis constructed with first $n$ principal components of the set of solutions $\{\tilde\phi_l\}_{l=1}^N$ if $N\geq n$, and returns $POD(\{\tilde\phi_l\}_{l=1}^N,N)$ if $N<n$. This function can be used to construct a POD basis of size up to $n$ with a set of solutions. The Greedy strategy we used is summarized in Algorithm~\ref{alg:greed}. The desired POD bases were constructed by running $POD(RB,n)$ on the output of this algorithm with $\alpha_1=[q_1,\kappa_1]=[15, 17.5]$, $n_2=10$, $N=20$, and $N_2=100$.

\subsection{Spherical Nuclear Density Functional Theory}
The four training points, $\phi_k$, used in Fig.~2 b) of the main text only varied the $\knm$ and $\lsym$ parameters, with the rest taken to be the standard UNEDF1 optimal parameters~\cite{kortelainen2011}. The four values of $\knm$ and $\lsym$, in MeV, are:
\begin{align*}
     \knm=200, \lsym=30;  
     \knm=200, \lsym=60; \\
     \knm=220, \lsym=30;
     \knm=220, \lsym=60.
\end{align*}

Table~\ref{tab:results SM} shows the parameter ranges used for the LHS for the DFT results in Table~I of the main text. Both the 500 testing points and the $N=20$ exact evaluations used to build the POD RB were independently drawn by LHS on these ranges.

\section{Glossary of terminology}


Table~\ref{tab:acronyms} shows the frequently used terms in this work (inspired by \cite{bedaque2021ai}).

\begin{table*}[h!]
 \caption{Glossary of acronyms
 }
    \label{tab:acronyms}
\begin{ruledtabular}
    \begin{tabular}{C{0.03\linewidth} p{0.15\linewidth}
    p{0.45\linewidth}C{0.15\linewidth}}
    
    Acronym & 
  \multicolumn{1}{c}{Name} & 
      \multicolumn{1}{c}{Brief Description} & Detailed Ref.\\ \hline\\[-5pt] 
      
       EC & Eigenvector Continuation   & Numerical method for approximating the ``trajectory" of an eigenvector associated with a parametrized operator as the corresponding parameters change. As shown in this article, it can be seen as a special case of the RBM. & \cite{frame2018eigenvector,sarkar2021convergence} \\
RBM & Reduced Basis Method  & Numerical method for solving parametrized differential equations efficiently by using a handful of previously computed solutions.  & Chapters 3 in \cite{hesthaven2016certified,quarteroni2015reduced}  \\
     SVD &  Singular Value Decomposition  & Matrix factorization algorithm key for many modern computational methods, including PCA and POD. & Chapter 1 in \cite{brunton2019data}, Chapter 2 in \cite{golub2013matrix} \\
  POD & Proper Orthogonal Decomposition   & SVD application to partial differential equations used to capture a low-dimensional representation of the corresponding dynamical system. In the context of RBMs it is used to construct small bases that capture a low-dimensional representation of a larger set of ``exact'' solutions.  & Sec. 3.3.1 in \cite{hesthaven2016certified} , Chapter 6 in \cite{quarteroni2015reduced}, Sec. 11.1 in \cite{brunton2019data} \\

        PCA & Principal Component Analysis  & SVD application where the variability of high-dimensional data is decomposed into its more statistically descriptive factors. & Chapter 1 in \cite{brunton2019data}, \cite{hotelling1933analysis,turk1991eigenfaces,jolliffe2016principal} \\
    
        LHS & Latin Hypercube Sampling  & Sampling technique for efficiently distributing points in $\mathbb{R}^n$.   & \cite{mckay2000comparison}\\
    - & Greedy Algorithm  & Algorithm that selects the locally optimal choice on each iteration. In the context of RBMs, it sequentially selects ``exact" solutions to train the emulator, usually by maximizing an estimated error. & Sec. 3.2.2 in \
    \cite{hesthaven2016certified}, Chapter 7 in \cite{quarteroni2015reduced}, \cite{Sarkar2021fpz}\\

    - & Lagrange Basis  & A reduced basis of size $n$ for the RBM that is built as a linear combination of only $n$ ``exact" solutions. & \cite{quarteroni2011certified} \\
     DFT & Density Functional Theory   & Mean-field approach to many-body quantum systems. & \cite{bender2003self,nucleardft} \\
     EDF  & Energy Density Functional  & The object that defines the interaction used in DFT.& \cite{bender2003self,nucleardft} \\

     \end{tabular}
 \end{ruledtabular}  
\end{table*}

\end{document}